\documentclass{appolb}
\usepackage{epsfig}

\begin{document}
\title{ Equation of state for isospin asymmetric nuclear matter using Lane potential }
{
\author{  D.N. Basu$^{1}$\thanks{E-mail:dnb@veccal.ernet.in}, P. Roy Chowdhury$^{2}$ and C. Samanta$^{2,3}$
\address{$^{1}$Variable Energy Cyclotron Centre, 1/AF Bidhan Nagar, Kolkata 700 064, India\\
$^{2}$Saha Institute of Nuclear Physics, 1/AF Bidhan Nagar, Kolkata 700 064, India\\
$^{3}$Physics Department, Virginia Commonwealth University, Richmond, VA 23284-2000, U.S.A.}
}
\vskip 0.2cm
\maketitle
\begin{abstract}

     A mean field calculation for obtaining the equation of state (EOS) for symmetric nuclear matter from a density dependent M3Y interaction supplemented by a zero-range potential is described. The energy per nucleon is minimized to obtain the ground state of symmetric nuclear matter. The saturation energy per nucleon used for nuclear matter calculations is determined from the co-efficient of the volume term of Bethe-Weizs\"acker mass formula which is evaluated by fitting the recent experimental and estimated atomic mass excesses from Audi-Wapstra-Thibault atomic mass table by minimizing the mean square deviation. The constants of density dependence of the effective  interaction are obtained by reproducing the saturation energy per nucleon and the saturation density of spin and isospin symmetric cold infinite nuclear matter. The EOS of symmetric nuclear matter, thus obtained, provide reasonably good estimate of nuclear incompressibility. Once the consants of density dependence are determined, EOS for asymmetric nuclear matter is calculated by adding to the isoscalar part, the isovector component of the M3Y interaction that do not contribute to the EOS of symmetric nuclear matter. These EOS are then used to calculate the pressure, the energy density and the velocity of sound in symmetric as well as isospin asymmetric nuclear matter. 
\vskip 0.5cm
\noindent 
Keywords : Asymmetric nuclear matter, Mass formula, Binding Energy, Atomic mass excess, Nuclear incompressibility, Nuclear symmetry energy.
 
\end{abstract}
\PACS{ 21.65.+f, 23.60.+e, 23.70.+j, 25.70.Bc, 21.30.Fe, 24.10.Ht }
\vskip 0.5cm





\section{Introduction}
\label{section1}

      The equation of state (EOS) of dense isospin asymmetric nuclear matter determines most of the gross properties of neutron stars and hence it is of considerable interest in astrophysics. Nuclear matter is an idealized system of nucleons interacting strongly through nuclear forces but without Coulomb forces and is translationally invariant with a fixed ratio of neutrons to protons. The nuclear EOS, which is the energy per nucleon E/A = $\epsilon$ of nuclear matter as a function of nucleonic density $\rho$, can be used to obtain the bulk properties of nuclear matter such as the nuclear incompressibility \cite{Ba80},\cite{Sa89} the energy density and the pressure needed for neutron star calculations and the velocity of sound in nuclear medium for predictions of shock wave generation and propagation. The EOS is also of fundamental importance in the theories of nucleus-nucleus collisions at energies where the nuclear incompressibility $K$ comes into play as well as in the theories of supernova explosions \cite{Be88}. In the present work we obtain an EOS for nuclear matter using the M3Y-Reid-Elliott effective interaction supplemented by a zero range pseudo-potential along with the density dependence. The density dependence parameters of the interaction are obtained by reproducing the saturation energy per nucleon and the saturation density of cold infinite spin and isospin symmetric nuclear matter (SNM). One of the density dependence parameter, which can be interpreted as the isospin averaged nucleon-nucleon interaction cross section in ground state symmetric nuclear medium, is also used to provide estimate for the nuclear mean free path. EOS for the isospin asymmetric nuclear matter is then calculated by adding to the isoscalar part, the Lane \cite{La62} or the isovector component \cite{Sa83} of the M3Y interaction that do not contribute to the EOS of SNM. These EOS are then used to calculate the pressure, the energy density and the velocity of sound in symmetric as well as isospin asymmetric nuclear matter and pure neutron matter (PNM). 

      The M3Y interaction was derived by fitting its matrix elements in an oscillator basis to those elements of the G-matrix obtained with the Reid-Elliott soft-core NN interaction. The ranges of the M3Y forces were chosen to ensure a long-range tail of the one-pion exchange potential as well as a short range repulsive part simulating the exchange of heavier mesons \cite{Be77}. The real part of the nuclear interaction potential obtained by folding in the density distribution functions of two interacting nuclei with the density dependent M3Y effective interaction supplemented by a zero-range pseudo-potential (DDM3Y) was shown to provide good descriptions for medium and high energy $\alpha$ and heavy ion elastic scatterings \cite{Sa79},\cite{Ko84},\cite{Gi87}. The zero-range pseudo-potential represented the single-nucleon exchange term while the density dependence accounted for the higher order exchange effects and the Pauli blocking effects. The real part of the protron-nucleus interaction potential obtained by folding in the density distribution function of interacting nucleus with the DDM3Y effective interaction is found to provide good descriptions of elastic and inelastic scatterings of high energy protons \cite{Gu05} and proton radioactivity \cite{BCS05}. Since the density dependence of the effective projectile-nucleon interaction was found to be fairly independent of the projectile \cite{Sr83}, as long as the projectile-nucleus interaction was amenable to a single-folding prescription, the density dependent effects on the nucleon-nucleon interaction were factorized into a target term times a projectile term and used successfully in case of $\alpha$ radioctivity of nuclei \cite{Ba03} including superheavies \cite{CSB06} and the cluster radioactivity \cite{Ba03}. 

\section{The density dependent effective nucleon-nucleon interaction : isoscalar and isovector components}
\label{section2}

      The central part of the effective interaction between two nucleons 1 and 2 can be written as  \cite{Sa79} 

\begin{equation}
 v_{12}(s) = v_{00}(s) + v_{01}(s) \tau_1.\tau_2 + v_{10}(s) \sigma_1.\sigma_2 
+ v_{11}(s) \sigma_1.\sigma_2~\tau_1.\tau_1
\label{seqn1}
\end{equation}   
\noindent 
where $\tau_1,\tau_2$ are the isospins and $\sigma_1,\sigma_2$ are the spins of nucleons 1,2. In case of SNM only the first term, the isoscalar term, contributes whereas for the isospin asymmetric-spin symmetric nuclear matter only first two terms, the isoscalar and the isovector (Lane) terms, contribute and for the spin-isospin asymmetric nuclear matter all the four terms of Eq.(1) contribute.  Considering only the isospin asymmetric-spin symmetric nuclear matter, the neutron-neutron, proton-proton, neutron-proton and proton-neutron interactions, {\it viz.} $v_{nn}, v_{pp}, v_{np}$ and $v_{pn}$, respectively, can be given by the following: 

\begin{equation}
 v_{nn} = v_{pp} =v_{00} + v_{01},~~~~v_{np} = v_{pn} = v_{00} - v_{01}
\label{seqn2}
\end{equation}   
\noindent 
The general expression for the density dependent effective NN interaction potential is written as \cite{BCS05} 

\begin{equation}
 v_{00}(s,\rho, \epsilon) = t_{00}^{M3Y}(s, \epsilon) g(\rho, \epsilon),~~~~v_{01}(s,\rho, \epsilon) = t_{01}^{M3Y}(s, \epsilon) g(\rho, \epsilon)
\label{seqn3}
\end{equation}   
\noindent
where the isoscalar $t_{00}^{M3Y}$ and the isovector $t_{01}^{M3Y}$ components of M3Y interaction potentials \cite{Sa79} supplemented by zero range potentials are given by the following: 

\begin{equation}
 t_{00}^{M3Y}(s, \epsilon) = 7999\frac{\exp( - 4s)}{4s} - 2134\frac{\exp( - 2.5s)}{2.5s} - 276 (1 - \alpha\epsilon)\delta(s)
\label{seqn4}
\end{equation}   
\noindent
and

\begin{equation}
  t_{01}^{M3Y}(s, \epsilon) =  -4886\frac{\exp( - 4s)}{4s} + 1176\frac{\exp( - 2.5s)}{2.5s} + 228 (1 - \alpha\epsilon)\delta(s)
\label{seqn5}
\end{equation}   
\noindent
where $s$ is the distance between two interacting nucleons and the energy dependence parameter $\alpha=0.005 MeV^{-1}$. The zero-range potentials of Eqs.(4,5) represent the single-nucleon exchange term. The density dependent part appearing in Eqs.(3) \cite{Ba05} has been taken to be of a general form

\begin{equation}
 g(\rho, \epsilon) = C (1 - \beta(\epsilon)\rho^n) 
\label{seqn6}
\end{equation}   
\noindent
which takes care of the higher order exchange effects and the Pauli blocking effects. This density dependence changes sign at high densities which is of crucial importance in fulfilling the saturation condition as well as giving different $K_0$ values with different values of $n$ for the nuclear EOS \cite{Ba05}. The value of the parameter $n=2/3$ was originally taken by Myers in the single folding calculation \cite{My73}. In fact $n=2/3$ also has a physical meaning because then $\beta$ can be interpreted as an 'in medium' effective nucleon-nucleon interaction cross-section $\sigma_0$ while the density dependent term represents interaction probability. This value of $\beta$ along with nucleonic density of infinite nuclear matter $\rho_0$ can also provide the nuclear mean free path $\lambda=1/(\rho_0 \sigma_0)$. Moreover, it also worked well in the single folding calculations for inelastic and elastic scatterings of high energy protons \cite{Gu05}, proton radioactivity \cite{BCS05} and in the double folding calculations with the factorized density dependence for $\alpha$ radioctivity of nuclei \cite{Ba03} including superheavies \cite{CSB06} and the cluster radioactivity \cite{Ba03}. 

\section{Symmetric and isospin asymmetric nuclear matter calculations}
\label{section3}

      The isospin asymmetry $X$ can be conveniently defined as 

\begin{equation}
 X = \frac{\rho_n-\rho_p}{\rho_n+\rho_p},~~~~\rho = \rho_n+\rho_p,
\label{seqn7}
\end{equation}   
\noindent
where $\rho_n$, $\rho_p$ and $\rho$ are the neutron, proton and nucleonic densities respectively. The asymmetry parameter $X$ can have values between -1  to +1, corresponding to pure proton matter and pure neutron matter respectively, while for SNM it becomes zero. For a single neutron interacting with rest of nuclear matter with isospin asymmetry $X$, the interaction energy per unit volume at $s$ is given by the following:

\begin{eqnarray}
 \rho_n v_{nn}(s)+\rho_p v_{np}(s)=&&\rho_n [v_{00}(s)+v_{01}(s)]+\rho_p[v_{00}(s)-v_{01}(s)] \nonumber \\
=&&[v_{00}(s)+v_{01}(s)X] \rho,
\label{seqn8}
\end{eqnarray}   
\noindent
while in case of a single proton interacting with rest of nuclear matter with isospin asymmetry $X$, the interaction energy per unit volume at $s$ is given by the following:

\begin{eqnarray}
 \rho_n v_{pn}(s)+\rho_p v_{pp}(s)=&&\rho_n [v_{00}(s)-v_{01}(s)]+\rho_p[v_{00}(s)+v_{01}(s)] \nonumber \\
=&&[v_{00}(s)-v_{01}(s)X] \rho,
\label{seqn9}
\end{eqnarray}   
\noindent
Summing the contributions for protons and neutrons and integrating over the entire volume of the infinite nuclear matter and multiplying by the factor $\frac{1}{2}$ to ignore the double counting in the process, the potential energy per nucleon $\epsilon_{pot}$ can be obtained by dividing the total potential energy by the total number of nucleons, 

\begin{equation}
 \epsilon_{pot}=\frac{g(\rho, \epsilon) \rho J_v}{2},
\label{seqn10}
\end{equation}   
\noindent
where

\begin{equation}
 J_v = J_{v00} + X^2 J_{v01} = \int \int \int [t_{00}^{M3Y}(s, \epsilon)+t_{01}^{M3Y}(s, \epsilon) X^2] d^3s . 
\label{seqn11}
\end{equation}   
\noindent

      Assuming interacting Fermi gas of neutrons and protons, the kinetic energy per nucleon $\epsilon_{kin}$ turns out to be  

\begin{equation}
 \epsilon_{kin} = [\frac{3\hbar^2k_F^2}{10m}] F(X), ~~~~F(X) = [\frac{(1+X)^{5/3} + (1-X)^{5/3}}{2}],
\label{seqn12}
\end{equation}   
\noindent
where $m$ is the nucleonic mass equal to 938.91897 $MeV/c^2$ and $k_F$, which becomes equal to Fermi momentum in case of the SNM, is given by the following:

\begin{equation}
 k_F^3 = 1.5\pi^2\rho,
\label{seqn13}
\end{equation}                                                                                                                                           
\noindent     

      The two parameters of Eq.(6), $C$ and $\beta$, are determined by reproducing the saturation conditions. It is worthwhile to mention here that due to attractive character of the M3Y forces the saturation condition for cold nuclear matter is not fulfilled. However, the realistic description of nuclear matter properties can be obtained with this density dependent M3Y effective interaction. Therefore, the density dependence parameters have been obtained by reproducing the saturation energy per nucleon and the saturation nucleonic density of the cold SNM.   

      The energy per nucleon $\epsilon=\epsilon_{kin}+\epsilon_{pot}$ obtained for the cold SNM for which $X=0$ is given by the following:

\begin{equation}
 \epsilon = [\frac{3\hbar^2k_F^2}{10m}] + \frac{g(\rho, \epsilon) \rho J_{v00}}{2}
\label{seqn14}
\end{equation}   
\noindent
where $J_{v00}$ represents the volume integral of the isoscalar part of the M3Y interaction supplemented by the zero-range potential having the form    
 
\begin{equation}
 J_{v00}(\epsilon) = \int \int \int t_{00}^{M3Y}(s, \epsilon) d^3s = 7999\frac{4\pi}{4^3} - 2134\frac{4\pi}{2.5^3} - 276 (1 - \alpha\epsilon)
\label{seqn15}
\end{equation}
\noindent
The Eq.(14) can be rewritten with the help of Eq.(6) as 

\begin{equation}
 \epsilon = [\frac{3\hbar^2k_F^2}{10m}] + [\frac{\rho J_{v00} C (1 - \beta\rho^n)}{2}]  
\label{seqn16}
\end{equation}
\noindent
and differentiated with respect to $\rho$ to yield equation  

\begin{equation}
 \frac{\partial\epsilon}{\partial\rho} = [\frac{\hbar^2k_F^2}{5m\rho}] + \frac{J_{v00} C}{2} [1 - (n+1)\beta\rho^n] 
\label{seqn17}
\end{equation}
\noindent
The equilibrium density of the cold SNM is determined from the saturation condition $\frac{\partial\epsilon}{\partial\rho} = 0$. Then Eq.(16) and Eq.(17) with the saturation condition can be solved simultaneously for fixed values of the saturation energy per nucleon $\epsilon_0$ and the saturation density $\rho_{0}$ of the cold SNM to obtain the values of the density dependence parameters $\beta$ and C. Density dependence parameters $\beta$ and C, thus obtained, can be given by the following:  

\begin{equation}
 \beta = \frac{[(1-p)\rho_{0}^{-n}]}{[(3n+1)-(n+1)p]},
\label{seqn18}
\end{equation} 
\noindent
where

\begin{equation}
 p = \frac{[10m\epsilon_0]}{[\hbar^2k_{F_0}^2]},
\label{seqn19}
\end{equation} 
\noindent
and 

\begin{equation}
 k_{F_0} = [1.5\pi^2\rho_0]^{1/3},
\label{seqn20}
\end{equation} 
\noindent

\begin{equation}
 C = -\frac{[2\hbar^2k_{F_0}^2] }{ 5mJ_{v00} \rho_0[1 - (n+1)\beta\rho_0^n]},
\label{seqn21}
\end{equation} 
\noindent
respectively. It is quite obvious that the density dependence parameter $\beta$ obtained by this method depends only on the saturation energy per nucleon $\epsilon_0$, the saturation density $\rho_{0}$ and the index $n$ of the density dependent part but not on the parameters of the M3Y interaction while the other density dependence parameter $C$ depends on the parameters of the M3Y interaction also through the volume integral $J_{v00}$. 

      The incompressibility $K_0$ of the cold SNM which is defined as   
  
\begin{equation}
 K_0 = k_F^2\frac{\partial^2\epsilon}{\partial{k_F^2}} \mid_{k_F=k_{F_0}} = 9\rho^2\frac{\partial^2\epsilon}{\partial\rho^2} \mid_{\rho=\rho_0}
\label{seqn22}
\end{equation}
\noindent
can be theoretically obtained using Eq.(13), Eq.(17) and Eq.(22) as

\begin{equation}
 K_{0} = [-(\frac{3\hbar^2k_{F_0}^2}{5m})] - [\frac{9 J_{v00} C n(n+1) \beta\rho_0^{n+1}}{2}]
\label{seqn23}
\end{equation} 
\noindent
Since the product $J_{v00} C$ appears in the above equation, a cursory glance reveals that the incompressibility $K_0$ depends only upon the saturation energy per nucleon $\epsilon_0$, the saturation density $\rho_{0}$ and the index $n$ of the density dependent part of the interaction but not on the parameters of the M3Y interaction.     
 
      The energy per nucleon for nuclear matter with isospin asymmetry $X$ can be rewritten as    

\begin{eqnarray}
 \epsilon = &&[\frac{3\hbar^2k_F^2}{10m}] F(X) + (\frac{\rho J_v C}{2}) (1 - \beta\rho^n) \nonumber \\
= &&[\frac{3\hbar^2k_F^2}{10m}] F(X) - [\frac{\rho}{\rho_{0}}] [\frac{J_v}{J_{v00}}] [\frac{\hbar^2k_{F_0}^2 (1 - \beta\rho^n)}{5m[1 - (n+1)\beta\rho_{0}^n]}]
\label{seqn24}
\end{eqnarray}
\noindent
where $J_v=J_{v00} + X^2 J_{v01}$ and $J_{v01}$ represents the volume integral of the isovector part of the M3Y interaction supplemented by the zero-range potential having the form    
 
\begin{equation}
 J_{v01}(\epsilon) = \int \int \int t_{01}^{M3Y}(s, \epsilon) d^3s = -4886\frac{4\pi}{4^3} + 1176\frac{4\pi}{2.5^3} + 228 (1 - \alpha\epsilon)
\label{seqn25}
\end{equation}
\noindent 
The pressure $P$ and the energy density $\varepsilon$ of nuclear matter with isospin asymmetry $X$ can be given by the following: 

\begin{equation}
 P = \rho^2 \frac{\partial\epsilon}{\partial\rho} = [\frac{\rho \hbar^2k_F^2}{5m}] F(X) + [\frac{\rho^2 J_v C}{2}] [1 - (n+1) \beta\rho^n],
\label{seqn26}
\end{equation} 
\noindent

\begin{equation}
 \varepsilon = \rho (\epsilon + m c^2) = \rho [(\frac{3\hbar^2k_F^2}{10m}) F(X) + (\frac{\rho J_v C}{2}) (1 - \beta\rho^n) + m c^2], 
\label{seqn27}
\end{equation} 
\noindent
respectively, and thus the velocity of sound $v_s$ in nuclear matter with isospin asymmetry $X$ is given by the following: 

\begin{equation}
 \frac{v_s}{c} = \sqrt{\frac{\partial P}{\partial\varepsilon}} =\sqrt{\frac{[2\rho\frac{\partial\epsilon}{\partial\rho}-\frac{\hbar^2k_F^2}{15m}F(X) - \frac{J_vCn(n+1)\beta \rho^{n+1}}{2}]} {[\epsilon + m c^2 + \rho\frac{\partial\epsilon}{\partial\rho}]}}
\label{seqn28}
\end{equation} 
\noindent
The incompressibilities for isospin asymmetric nuclear matter are evaluated at saturation densities $\rho_s$ with the condition $\frac{\partial\epsilon}{\partial\rho}=0$ which corresponds to vanishing pressure. The incompressibility $K$ for isospin asymmetric nuclear matter is therefore expressed as the following:  

\begin{equation}
 K_{0} = [-(\frac{3\hbar^2k_F^2}{5m})] F(X) - [\frac{9 J_v C n(n+1) \beta\rho_s^{n+1}}{2}]
\label{seqn29}
\end{equation} 
\noindent
where $k_F$ is now evaluated at saturation density $\rho_s$ using Eq.(13) and $J_v=J_{v00} + X^2 J_{v01}$.

\section{Calculations of energy per nucleon, pressure, energy density and velocity of sound for symmetric nuclear matter and neutron matter}
\label{section4}

      The calculations have been performed using the values of the saturation density $\rho_0=0.1533 fm^{-3}$ \cite{Ba90} and the saturation energy per nucleon $\epsilon_0=-15.26 MeV$ \cite{Ch05} for the SNM obtained from the co-efficient of the volume term of Bethe-Weizs\"acker mass formula which is evaluated by fitting the recent experimental and estimated atomic mass excesses from Audi-Wapstra-Thibault atomic mass table \cite{Au03} by minimizing the mean square deviation. For a fixed value of $\beta$, the parameters $\alpha$ and $C$ can have any possible simultaneous values as determined from SNM. Using the usual value of $\alpha=0.005 MeV^{-1}$ for the parameter of energy dependence of the zero range potential, the values obtained for the density dependence parameters $C$ and $\beta$ are presented in Table-1 for different values of the parameter $n$ along with the corresponding values of the incompressibility $K_0$. Smaller $n$ values predict softer EOS while higher values predict stiffer EOS. The form of $C(1-\beta\rho^n)$ with $n=2/3$ for the density dependence which is identical to that used for explaining the elastic and inelastic scattering \cite{Gu05} of protons and the proton \cite{BCS05}, $\alpha$ \cite{Ba03},\cite{CSB06}, cluster radioactivity phenomena \cite{Ba03} also agrees well with recent theoretical \cite{Sa99} and experimental \cite{Sc96} results for the nuclear incompressibility. 

\begin{table}[htbp]
\caption{Incompressiility of SNM for different values of n using the usual value of energy dependence  parameter $\alpha=0.005 MeV^{-1}$ and using values of saturation density $\rho_0=0.1533 fm^{-3}$ and saturation energy per nucleon $\epsilon_0=-15.26 MeV$.}
\centering
\begin{tabular}{|c|c|c|c|}
\hline
$n$&$\beta$&C & $K_{0}$      \\ \hline
 &$fm^{3n}$&          &$MeV$    \\ \hline

 1/6&1.074&5.05&192.5 \\  \hline
 1/3&1.208&3.07&226.1 \\  \hline
 2/3&1.668&2.07&293.4 \\  \hline
 1&2.472 &1.74 &360.6 \\  \hline
 2&9.947 &1.41 &562.4 \\  \hline
  
\end{tabular} 
\end{table}
\nopagebreak

      In Table-2 incompressiility of isospin asymmetric nuclear matter as a function of the isospin asymmetry parameter $X$, using the usual value of n=$2/3$ and energy dependence parameter $\alpha=0.005 MeV^{-1}$, is provided. 

\begin{table}[htbp]
\caption{Incompressiility of isospin asymmetric nuclear matter using the usual value of n=$2/3$ and energy dependence  parameter $\alpha=0.005 MeV^{-1}$.}
\centering
\begin{tabular}{|c|c|c|}
\hline
$X$&$\rho_s$& $K_{0}$      \\ \hline
 & $fm^{-3}$ &$MeV$    \\ \hline

 0.0&0.1533&293.4 \\ \hline
 0.1&0.1526&288.8 \\ \hline
 0.2&0.1503&275.3 \\ \hline
 0.3&0.1464&252.9 \\ \hline
 0.4&0.1403&221.7 \\ \hline
 0.5&0.1315&182.0 \\ \hline
  
\end{tabular} 
\end{table}
\nopagebreak

      In Tables-3-5 the theoretical estimates of the pressure $P$ and velocity of sound $v_s$ of SNM are listed as functions of nucleonic density $\rho$ and energy density $\varepsilon$ using the usual value of 0.005 $MeV^{-1}$ for the parameter $\alpha$ of energy dependence, given in Eqs.(4,5), of the zero range potential and also the standard value of the parameter $n=2/3$. As for any other non-relativistic EOS, present EOS also suffers from superluminosity at very high densities. According to present calculations the velocity of sound becomes imaginary for $\rho\le 0.1fm^{-3}$ and exceeds the velocity of light c at $\rho \ge5.3\rho_0$ and the EOS obtained using $v_{14}+TNI$ \cite{Fr81} also resulted in sound velocity becoming imaginary at same nuclear density and  superluminous at about the same nuclear density. But in contrast, the incompressibility $K_0$ of infinite SNM for the $v_{14}+TNI$ was chosen to be 240 MeV while that by the present theoretical estimate is about 290 MeV which is in excellent agreement with the experimental value of $K_0=300\pm25$ MeV obtained from the giant monopole resonance (GMR) \cite{Sh88} and with the the recent experimental determination of $K_0$ based upon the production of hard photons in heavy ion collisions which led to the experimental estimate of $K_0=290\pm50$ MeV \cite{Sc96}.

\begin{table}[htbp]
\caption{$\frac{\rm{Energy}}{\rm{nucleon}}~\epsilon$, pressure $P$, energy density $\varepsilon$ and velocity of sound $v_s$ as functions of nucleonic density $\rho$ for SNM.}
\centering
\begin{tabular}{|c|c|c|c|c|c|}
\hline
$\rho$&$\rho/\rho_{0}$&$\epsilon$&P&$\varepsilon $&$v_s$  \\ \hline
$fm^{-3}$& &$MeV$   &$MeV fm^{-3}$&$MeV fm^{-3}$&in units of c       \\ \hline

     .01&  .6523E-01& -.7537E+00& -.1677E-01&  .9382E+01&  .0000E+00\\
     .02&  .1305E+00& -.2526E+01& -.7232E-01&  .1873E+02&  .0000E+00\\
     .03&  .1957E+00& -.4312E+01& -.1574E+00&  .2804E+02&  .0000E+00\\
     .04&  .2609E+00& -.6007E+01& -.2617E+00&  .3732E+02&  .0000E+00\\
     .05&  .3262E+00& -.7576E+01& -.3752E+00&  .4657E+02&  .0000E+00\\
     .06&  .3914E+00& -.9006E+01& -.4885E+00&  .5579E+02&  .0000E+00\\
     .07&  .4566E+00& -.1029E+02& -.5926E+00&  .6500E+02&  .0000E+00\\
     .08&  .5219E+00& -.1142E+02& -.6786E+00&  .7420E+02&  .0000E+00\\
     .09&  .5871E+00& -.1241E+02& -.7382E+00&  .8339E+02&  .0000E+00\\
     .10&  .6523E+00& -.1325E+02& -.7633E+00&  .9257E+02&  .0000E+00\\
     .11&  .7175E+00& -.1394E+02& -.7460E+00&  .1017E+03&  .6683E-01\\
     .12&  .7828E+00& -.1448E+02& -.6787E+00&  .1109E+03&  .1016E+00\\
     .13&  .8480E+00& -.1488E+02& -.5540E+00&  .1201E+03&  .1302E+00\\
     .14&  .9132E+00& -.1514E+02& -.3645E+00&  .1293E+03&  .1560E+00\\
     .15&  .9785E+00& -.1525E+02& -.1032E+00&  .1385E+03&  .1802E+00\\
     .16&  .1044E+01& -.1523E+02&  .2369E+00&  .1478E+03&  .2031E+00\\
     .17&  .1109E+01& -.1507E+02&  .6627E+00&  .1571E+03&  .2253E+00\\
     .18&  .1174E+01& -.1477E+02&  .1181E+01&  .1663E+03&  .2467E+00\\
     .19&  .1239E+01& -.1434E+02&  .1798E+01&  .1757E+03&  .2675E+00\\
     .20&  .1305E+01& -.1378E+02&  .2520E+01&  .1850E+03&  .2879E+00\\
     .21&  .1370E+01& -.1308E+02&  .3354E+01&  .1944E+03&  .3077E+00\\
     .22&  .1435E+01& -.1226E+02&  .4306E+01&  .2039E+03&  .3272E+00\\
     .23&  .1500E+01& -.1130E+02&  .5382E+01&  .2134E+03&  .3463E+00\\
     .24&  .1566E+01& -.1022E+02&  .6588E+01&  .2229E+03&  .3649E+00\\
     .25&  .1631E+01& -.9014E+01&  .7931E+01&  .2325E+03&  .3833E+00\\
     .26&  .1696E+01& -.7683E+01&  .9416E+01&  .2421E+03&  .4013E+00\\
     .27&  .1761E+01& -.6229E+01&  .1105E+02&  .2518E+03&  .4189E+00\\
     .28&  .1826E+01& -.4652E+01&  .1284E+02&  .2616E+03&  .4363E+00\\
     .29&  .1892E+01& -.2955E+01&  .1478E+02&  .2714E+03&  .4533E+00\\
     .30&  .1957E+01& -.1138E+01&  .1689E+02&  .2813E+03&  .4700E+00\\
     .31&  .2022E+01&  .7985E+00&  .1917E+02&  .2913E+03&  .4864E+00\\
     .32&  .2087E+01&  .2852E+01&  .2163E+02&  .3014E+03&  .5024E+00\\
     .33&  .2153E+01&  .5023E+01&  .2427E+02&  .3115E+03&  .5182E+00\\
     .34&  .2218E+01&  .7310E+01&  .2710E+02&  .3217E+03&  .5337E+00\\
     .35&  .2283E+01&  .9711E+01&  .3012E+02&  .3320E+03&  .5489E+00\\
     .36&  .2348E+01&  .1223E+02&  .3334E+02&  .3424E+03&  .5638E+00\\
     .37&  .2414E+01&  .1486E+02&  .3676E+02&  .3529E+03&  .5785E+00\\
     .38&  .2479E+01&  .1760E+02&  .4039E+02&  .3635E+03&  .5928E+00\\
     .39&  .2544E+01&  .2045E+02&  .4423E+02&  .3742E+03&  .6069E+00\\
     .40&  .2609E+01&  .2341E+02&  .4829E+02&  .3849E+03&  .6207E+00\\ \hline
\end{tabular} 
\end{table}
\nopagebreak

\begin{table}[htbp]
\caption{$\frac{\rm{Energy}}{\rm{nucleon}}~\epsilon$, pressure $P$, energy density $\varepsilon$ and velocity of sound $v_s$ as functions of nucleonic density $\rho$ for SNM.}
\centering
\begin{tabular}{|c|c|c|c|c|c|}
\hline
$\rho$&$\rho/\rho_{0}$&$\epsilon$&P&$\varepsilon $&$v_s$  \\ \hline
$fm^{-3}$& &$MeV$   &$MeV fm^{-3}$&$MeV fm^{-3}$&in units of c       \\ \hline
     .41&  .2674E+01&  .2648E+02&  .5257E+02&  .3958E+03&  .6342E+00\\
     .42&  .2740E+01&  .2967E+02&  .5709E+02&  .4068E+03&  .6474E+00\\
     .43&  .2805E+01&  .3296E+02&  .6184E+02&  .4179E+03&  .6604E+00\\
     .44&  .2870E+01&  .3636E+02&  .6682E+02&  .4291E+03&  .6731E+00\\
     .45&  .2935E+01&  .3986E+02&  .7205E+02&  .4405E+03&  .6856E+00\\
     .46&  .3001E+01&  .4347E+02&  .7753E+02&  .4519E+03&  .6978E+00\\
     .47&  .3066E+01&  .4719E+02&  .8326E+02&  .4635E+03&  .7098E+00\\
     .48&  .3131E+01&  .5101E+02&  .8925E+02&  .4752E+03&  .7215E+00\\
     .49&  .3196E+01&  .5493E+02&  .9550E+02&  .4870E+03&  .7330E+00\\
     .50&  .3262E+01&  .5896E+02&  .1020E+03&  .4989E+03&  .7442E+00\\
     .51&  .3327E+01&  .6310E+02&  .1088E+03&  .5110E+03&  .7552E+00\\
     .52&  .3392E+01&  .6733E+02&  .1159E+03&  .5233E+03&  .7660E+00\\
     .53&  .3457E+01&  .7167E+02&  .1232E+03&  .5356E+03&  .7765E+00\\
     .54&  .3523E+01&  .7611E+02&  .1309E+03&  .5481E+03&  .7868E+00\\
     .55&  .3588E+01&  .8065E+02&  .1388E+03&  .5608E+03&  .7969E+00\\
     .56&  .3653E+01&  .8528E+02&  .1470E+03&  .5736E+03&  .8068E+00\\
     .57&  .3718E+01&  .9002E+02&  .1556E+03&  .5865E+03&  .8165E+00\\
     .58&  .3783E+01&  .9486E+02&  .1644E+03&  .5996E+03&  .8259E+00\\
     .59&  .3849E+01&  .9980E+02&  .1735E+03&  .6128E+03&  .8352E+00\\
     .60&  .3914E+01&  .1048E+03&  .1830E+03&  .6262E+03&  .8443E+00\\
     .61&  .3979E+01&  .1100E+03&  .1928E+03&  .6398E+03&  .8531E+00\\
     .62&  .4044E+01&  .1152E+03&  .2029E+03&  .6535E+03&  .8618E+00\\
     .63&  .4110E+01&  .1205E+03&  .2133E+03&  .6674E+03&  .8703E+00\\
     .64&  .4175E+01&  .1259E+03&  .2240E+03&  .6815E+03&  .8786E+00\\
     .65&  .4240E+01&  .1315E+03&  .2351E+03&  .6957E+03&  .8867E+00\\
     .66&  .4305E+01&  .1371E+03&  .2466E+03&  .7102E+03&  .8947E+00\\
     .67&  .4371E+01&  .1428E+03&  .2583E+03&  .7247E+03&  .9025E+00\\
     .68&  .4436E+01&  .1486E+03&  .2705E+03&  .7395E+03&  .9101E+00\\
     .69&  .4501E+01&  .1545E+03&  .2830E+03&  .7544E+03&  .9175E+00\\
     .70&  .4566E+01&  .1605E+03&  .2958E+03&  .7696E+03&  .9248E+00\\
     .71&  .4631E+01&  .1665E+03&  .3090E+03&  .7849E+03&  .9319E+00\\
     .72&  .4697E+01&  .1727E+03&  .3225E+03&  .8004E+03&  .9389E+00\\
     .73&  .4762E+01&  .1790E+03&  .3365E+03&  .8161E+03&  .9457E+00\\
     .74&  .4827E+01&  .1854E+03&  .3508E+03&  .8320E+03&  .9524E+00\\
     .75&  .4892E+01&  .1918E+03&  .3655E+03&  .8480E+03&  .9589E+00\\
     .76&  .4958E+01&  .1983E+03&  .3805E+03&  .8643E+03&  .9653E+00\\
     .77&  .5023E+01&  .2050E+03&  .3960E+03&  .8808E+03&  .9715E+00\\
     .78&  .5088E+01&  .2117E+03&  .4118E+03&  .8975E+03&  .9776E+00\\
     .79&  .5153E+01&  .2185E+03&  .4281E+03&  .9144E+03&  .9836E+00\\
     .80&  .5219E+01&  .2254E+03&  .4447E+03&  .9315E+03&  .9895E+00\\ \hline
\end{tabular} 
\end{table}
\nopagebreak

\begin{table}[htbp]
\caption{$\frac{\rm{Energy}}{\rm{nucleon}}~\epsilon$, pressure $P$, energy density $\varepsilon$ and velocity of sound $v_s$ as functions of nucleonic density $\rho$ for SNM.}
\centering
\begin{tabular}{|c|c|c|c|c|c|}
\hline
$\rho$&$\rho/\rho_{0}$&$\epsilon$&P&$\varepsilon $&$v_s$  \\ \hline
$fm^{-3}$& &$MeV$   &$MeV fm^{-3}$&$MeV fm^{-3}$&in units of c       \\ \hline
     .81&  .5284E+01&  .2324E+03&  .4618E+03&  .9488E+03&  .9952E+00\\
     .82&  .5349E+01&  .2395E+03&  .4792E+03&  .9663E+03&  .1001E+01\\
     .83&  .5414E+01&  .2467E+03&  .4971E+03&  .9840E+03&  .1006E+01\\
     .84&  .5479E+01&  .2539E+03&  .5154E+03&  .1002E+04&  .1012E+01\\
     .85&  .5545E+01&  .2613E+03&  .5341E+03&  .1020E+04&  .1017E+01\\
     .86&  .5610E+01&  .2687E+03&  .5532E+03&  .1039E+04&  .1022E+01\\
     .87&  .5675E+01&  .2762E+03&  .5727E+03&  .1057E+04&  .1027E+01\\
     .88&  .5740E+01&  .2838E+03&  .5927E+03&  .1076E+04&  .1032E+01\\
     .89&  .5806E+01&  .2915E+03&  .6131E+03&  .1095E+04&  .1037E+01\\
     .90&  .5871E+01&  .2993E+03&  .6340E+03&  .1114E+04&  .1042E+01\\
     .91&  .5936E+01&  .3072E+03&  .6553E+03&  .1134E+04&  .1046E+01\\
     .92&  .6001E+01&  .3152E+03&  .6770E+03&  .1154E+04&  .1051E+01\\
     .93&  .6067E+01&  .3232E+03&  .6992E+03&  .1174E+04&  .1055E+01\\
     .94&  .6132E+01&  .3313E+03&  .7219E+03&  .1194E+04&  .1059E+01\\
     .95&  .6197E+01&  .3395E+03&  .7450E+03&  .1215E+04&  .1064E+01\\
     .96&  .6262E+01&  .3478E+03&  .7685E+03&  .1235E+04&  .1068E+01\\
     .97&  .6327E+01&  .3562E+03&  .7926E+03&  .1256E+04&  .1072E+01\\
     .98&  .6393E+01&  .3647E+03&  .8171E+03&  .1278E+04&  .1076E+01\\
     .99&  .6458E+01&  .3732E+03&  .8420E+03&  .1299E+04&  .1080E+01\\
    1.00&  .6523E+01&  .3819E+03&  .8675E+03&  .1321E+04&  .1084E+01\\ \hline
\end{tabular} 
\end{table}
\nopagebreak

      In Tables-6-8 the theoretical estimates of the pressure $P$ and velocity of sound $v_s$ in case of PNM are listed as functions of nucleonic density $\rho$ and energy density $\varepsilon$ using the usual value of 0.005 $MeV^{-1}$ for the parameter of energy dependence, given in Eqs.(4,5), of the zero range potential and also the standard value of the parameter $n=2/3$.

\begin{table}[htbp]
\caption{$\frac{\rm{Energy}}{\rm{nucleon}}~\epsilon$, pressure $P$, energy density $\varepsilon$ and velocity of sound $v_s$ as functions of nucleonic density $\rho$ for PNM.}
\centering
\begin{tabular}{|c|c|c|c|c|c|}
\hline
$\rho$&$\rho/\rho_{0}$&$\epsilon$&P&$\varepsilon $&$v_s$  \\ \hline
$fm^{-3}$& &$MeV$   &$MeV fm^{-3}$&$MeV fm^{-3}$&in units of c       \\ \hline

     .01&  .6523E-01&  .3509E+01&  .1780E-01&  .9424E+01&  .5166E-01\\
     .02&  .1305E+00&  .4937E+01&  .4742E-01&  .1888E+02&  .5986E-01\\
     .03&  .1957E+00&  .5992E+01&  .8594E-01&  .2835E+02&  .6774E-01\\
     .04&  .2609E+00&  .6886E+01&  .1353E+00&  .3783E+02&  .7657E-01\\
     .05&  .3262E+00&  .7702E+01&  .1983E+00&  .4733E+02&  .8637E-01\\
     .06&  .3914E+00&  .8483E+01&  .2782E+00&  .5684E+02&  .9693E-01\\
     .07&  .4566E+00&  .9254E+01&  .3783E+00&  .6637E+02&  .1081E+00\\
     .08&  .5219E+00&  .1003E+02&  .5020E+00&  .7592E+02&  .1196E+00\\
     .09&  .5871E+00&  .1083E+02&  .6526E+00&  .8548E+02&  .1314E+00\\
     .10&  .6523E+00&  .1164E+02&  .8334E+00&  .9506E+02&  .1433E+00\\
     .11&  .7175E+00&  .1249E+02&  .1048E+01&  .1047E+03&  .1554E+00\\
     .12&  .7828E+00&  .1338E+02&  .1299E+01&  .1143E+03&  .1676E+00\\
     .13&  .8480E+00&  .1430E+02&  .1590E+01&  .1239E+03&  .1797E+00\\
     .14&  .9132E+00&  .1526E+02&  .1924E+01&  .1336E+03&  .1919E+00\\
     .15&  .9785E+00&  .1626E+02&  .2304E+01&  .1433E+03&  .2041E+00\\
     .16&  .1044E+01&  .1731E+02&  .2733E+01&  .1530E+03&  .2162E+00\\
     .17&  .1109E+01&  .1840E+02&  .3214E+01&  .1627E+03&  .2282E+00\\
     .18&  .1174E+01&  .1953E+02&  .3751E+01&  .1725E+03&  .2402E+00\\
     .19&  .1239E+01&  .2071E+02&  .4345E+01&  .1823E+03&  .2521E+00\\
     .20&  .1305E+01&  .2194E+02&  .5001E+01&  .1922E+03&  .2640E+00\\
     .21&  .1370E+01&  .2321E+02&  .5720E+01&  .2020E+03&  .2757E+00\\
     .22&  .1435E+01&  .2453E+02&  .6506E+01&  .2120E+03&  .2874E+00\\
     .23&  .1500E+01&  .2590E+02&  .7361E+01&  .2219E+03&  .2989E+00\\
     .24&  .1566E+01&  .2732E+02&  .8288E+01&  .2319E+03&  .3104E+00\\
     .25&  .1631E+01&  .2878E+02&  .9290E+01&  .2419E+03&  .3218E+00\\
     .26&  .1696E+01&  .3029E+02&  .1037E+02&  .2520E+03&  .3330E+00\\
     .27&  .1761E+01&  .3185E+02&  .1153E+02&  .2621E+03&  .3442E+00\\
     .28&  .1826E+01&  .3345E+02&  .1277E+02&  .2723E+03&  .3553E+00\\
     .29&  .1892E+01&  .3511E+02&  .1410E+02&  .2825E+03&  .3662E+00\\
     .30&  .1957E+01&  .3681E+02&  .1552E+02&  .2927E+03&  .3771E+00\\
     .31&  .2022E+01&  .3855E+02&  .1702E+02&  .3030E+03&  .3878E+00\\
     .32&  .2087E+01&  .4035E+02&  .1862E+02&  .3134E+03&  .3984E+00\\
     .33&  .2153E+01&  .4219E+02&  .2032E+02&  .3238E+03&  .4090E+00\\
     .34&  .2218E+01&  .4408E+02&  .2211E+02&  .3342E+03&  .4194E+00\\
     .35&  .2283E+01&  .4602E+02&  .2401E+02&  .3447E+03&  .4297E+00\\
     .36&  .2348E+01&  .4800E+02&  .2600E+02&  .3553E+03&  .4399E+00\\
     .37&  .2414E+01&  .5003E+02&  .2810E+02&  .3659E+03&  .4499E+00\\
     .38&  .2479E+01&  .5211E+02&  .3031E+02&  .3766E+03&  .4599E+00\\
     .39&  .2544E+01&  .5423E+02&  .3264E+02&  .3873E+03&  .4698E+00\\
     .40&  .2609E+01&  .5640E+02&  .3507E+02&  .3981E+03&  .4795E+00\\ \hline
\end{tabular} 
\end{table}
\nopagebreak

\begin{table}[htbp]
\caption{$\frac{\rm{Energy}}{\rm{nucleon}}~\epsilon$, pressure $P$, energy density $\varepsilon$ and velocity of sound $v_s$ as functions of nucleonic density $\rho$ for PNM.}
\centering
\begin{tabular}{|c|c|c|c|c|c|}
\hline
$\rho$&$\rho/\rho_{0}$&$\epsilon$&P&$\varepsilon $&$v_s$  \\ \hline
$fm^{-3}$& &$MeV$   &$MeV fm^{-3}$&$MeV fm^{-3}$&in units of c       \\ \hline
     .41&  .2674E+01&  .5861E+02&  .3762E+02&  .4090E+03&  .4891E+00\\
     .42&  .2740E+01&  .6087E+02&  .4028E+02&  .4199E+03&  .4987E+00\\
     .43&  .2805E+01&  .6318E+02&  .4307E+02&  .4309E+03&  .5081E+00\\
     .44&  .2870E+01&  .6553E+02&  .4597E+02&  .4420E+03&  .5174E+00\\
     .45&  .2935E+01&  .6793E+02&  .4900E+02&  .4531E+03&  .5266E+00\\
     .46&  .3001E+01&  .7037E+02&  .5216E+02&  .4643E+03&  .5357E+00\\
     .47&  .3066E+01&  .7286E+02&  .5545E+02&  .4755E+03&  .5447E+00\\
     .48&  .3131E+01&  .7539E+02&  .5887E+02&  .4869E+03&  .5535E+00\\
     .49&  .3196E+01&  .7797E+02&  .6242E+02&  .4983E+03&  .5623E+00\\
     .50&  .3262E+01&  .8059E+02&  .6610E+02&  .5098E+03&  .5709E+00\\
     .51&  .3327E+01&  .8326E+02&  .6992E+02&  .5213E+03&  .5795E+00\\
     .52&  .3392E+01&  .8597E+02&  .7389E+02&  .5329E+03&  .5879E+00\\
     .53&  .3457E+01&  .8872E+02&  .7799E+02&  .5446E+03&  .5963E+00\\
     .54&  .3523E+01&  .9152E+02&  .8224E+02&  .5564E+03&  .6045E+00\\
     .55&  .3588E+01&  .9436E+02&  .8664E+02&  .5683E+03&  .6127E+00\\
     .56&  .3653E+01&  .9725E+02&  .9118E+02&  .5803E+03&  .6207E+00\\
     .57&  .3718E+01&  .1002E+03&  .9588E+02&  .5923E+03&  .6286E+00\\
     .58&  .3783E+01&  .1031E+03&  .1007E+03&  .6044E+03&  .6365E+00\\
     .59&  .3849E+01&  .1062E+03&  .1057E+03&  .6166E+03&  .6442E+00\\
     .60&  .3914E+01&  .1092E+03&  .1109E+03&  .6289E+03&  .6518E+00\\
     .61&  .3979E+01&  .1123E+03&  .1162E+03&  .6413E+03&  .6594E+00\\
     .62&  .4044E+01&  .1155E+03&  .1217E+03&  .6537E+03&  .6668E+00\\
     .63&  .4110E+01&  .1187E+03&  .1273E+03&  .6663E+03&  .6741E+00\\
     .64&  .4175E+01&  .1219E+03&  .1331E+03&  .6789E+03&  .6814E+00\\
     .65&  .4240E+01&  .1252E+03&  .1391E+03&  .6916E+03&  .6885E+00\\
     .66&  .4305E+01&  .1285E+03&  .1453E+03&  .7045E+03&  .6956E+00\\
     .67&  .4371E+01&  .1318E+03&  .1516E+03&  .7174E+03&  .7026E+00\\
     .68&  .4436E+01&  .1352E+03&  .1581E+03&  .7304E+03&  .7094E+00\\
     .69&  .4501E+01&  .1387E+03&  .1647E+03&  .7435E+03&  .7162E+00\\
     .70&  .4566E+01&  .1421E+03&  .1716E+03&  .7567E+03&  .7229E+00\\
     .71&  .4631E+01&  .1457E+03&  .1786E+03&  .7701E+03&  .7295E+00\\
     .72&  .4697E+01&  .1492E+03&  .1858E+03&  .7835E+03&  .7360E+00\\
     .73&  .4762E+01&  .1528E+03&  .1932E+03&  .7970E+03&  .7425E+00\\
     .74&  .4827E+01&  .1565E+03&  .2007E+03&  .8106E+03&  .7488E+00\\
     .75&  .4892E+01&  .1602E+03&  .2085E+03&  .8243E+03&  .7551E+00\\
     .76&  .4958E+01&  .1639E+03&  .2164E+03&  .8381E+03&  .7613E+00\\
     .77&  .5023E+01&  .1677E+03&  .2246E+03&  .8521E+03&  .7674E+00\\
     .78&  .5088E+01&  .1715E+03&  .2329E+03&  .8661E+03&  .7734E+00\\
     .79&  .5153E+01&  .1753E+03&  .2414E+03&  .8802E+03&  .7793E+00\\
     .80&  .5219E+01&  .1792E+03&  .2502E+03&  .8945E+03&  .7852E+00\\ \hline
\end{tabular} 
\end{table}
\nopagebreak

\begin{table}[htbp]
\caption{$\frac{\rm{Energy}}{\rm{nucleon}}~\epsilon$, pressure $P$, energy density $\varepsilon$ and velocity of sound $v_s$ as functions of nucleonic density $\rho$ for PNM.}
\centering
\begin{tabular}{|c|c|c|c|c|c|}
\hline
$\rho$&$\rho/\rho_{0}$&$\epsilon$&P&$\varepsilon $&$v_s$  \\ \hline
$fm^{-3}$& &$MeV$   &$MeV fm^{-3}$&$MeV fm^{-3}$&in units of c       \\ \hline
     .81&  .5284E+01&  .1831E+03&  .2591E+03&  .9089E+03&  .7909E+00\\
     .82&  .5349E+01&  .1871E+03&  .2682E+03&  .9233E+03&  .7966E+00\\
     .83&  .5414E+01&  .1911E+03&  .2775E+03&  .9379E+03&  .8023E+00\\
     .84&  .5479E+01&  .1952E+03&  .2871E+03&  .9526E+03&  .8078E+00\\
     .85&  .5545E+01&  .1992E+03&  .2968E+03&  .9674E+03&  .8133E+00\\
     .86&  .5610E+01&  .2034E+03&  .3067E+03&  .9824E+03&  .8187E+00\\
     .87&  .5675E+01&  .2075E+03&  .3169E+03&  .9974E+03&  .8240E+00\\
     .88&  .5740E+01&  .2117E+03&  .3272E+03&  .1013E+04&  .8293E+00\\
     .89&  .5806E+01&  .2160E+03&  .3378E+03&  .1028E+04&  .8345E+00\\
     .90&  .5871E+01&  .2203E+03&  .3486E+03&  .1043E+04&  .8396E+00\\
     .91&  .5936E+01&  .2246E+03&  .3596E+03&  .1059E+04&  .8446E+00\\
     .92&  .6001E+01&  .2290E+03&  .3709E+03&  .1074E+04&  .8496E+00\\
     .93&  .6067E+01&  .2334E+03&  .3823E+03&  .1090E+04&  .8545E+00\\
     .94&  .6132E+01&  .2378E+03&  .3940E+03&  .1106E+04&  .8594E+00\\
     .95&  .6197E+01&  .2423E+03&  .4059E+03&  .1122E+04&  .8641E+00\\
     .96&  .6262E+01&  .2468E+03&  .4180E+03&  .1138E+04&  .8689E+00\\
     .97&  .6327E+01&  .2514E+03&  .4304E+03&  .1155E+04&  .8735E+00\\
     .98&  .6393E+01&  .2559E+03&  .4429E+03&  .1171E+04&  .8781E+00\\
     .99&  .6458E+01&  .2606E+03&  .4557E+03&  .1187E+04&  .8826E+00\\
    1.00&  .6523E+01&  .2652E+03&  .4688E+03&  .1204E+04&  .8871E+00\\ \hline
\end{tabular} 
\end{table}
\nopagebreak

      In Fig.-1 the energy per nucleon $\epsilon$ of SNM and PNM are plotted as a function of $\rho$. The continuous lines represent the curves for the present calculations using saturation energy per nucleon of -15.26 MeV whereas the dotted lines represent the same using $v_{14}+TNI$ interaction \cite{Fr81} and the dash-dotted lines represent the same for the A18 model using variational chain summation (VCS) \cite{Ak98} for the SNM and PNM. The minimum of the energy per nucleon equaling the saturation energy of -15.26 MeV for the present calculations occurs precisely at the saturation density $\rho_0=0.1533 fm^{-3}$ while that for the A18(VCS) model occurs around $\rho=0.28 fm^{-3}$ with a saturation energy of about -17.3 MeV. 

\begin{figure}[htbp]
\eject\centerline{\epsfig{file=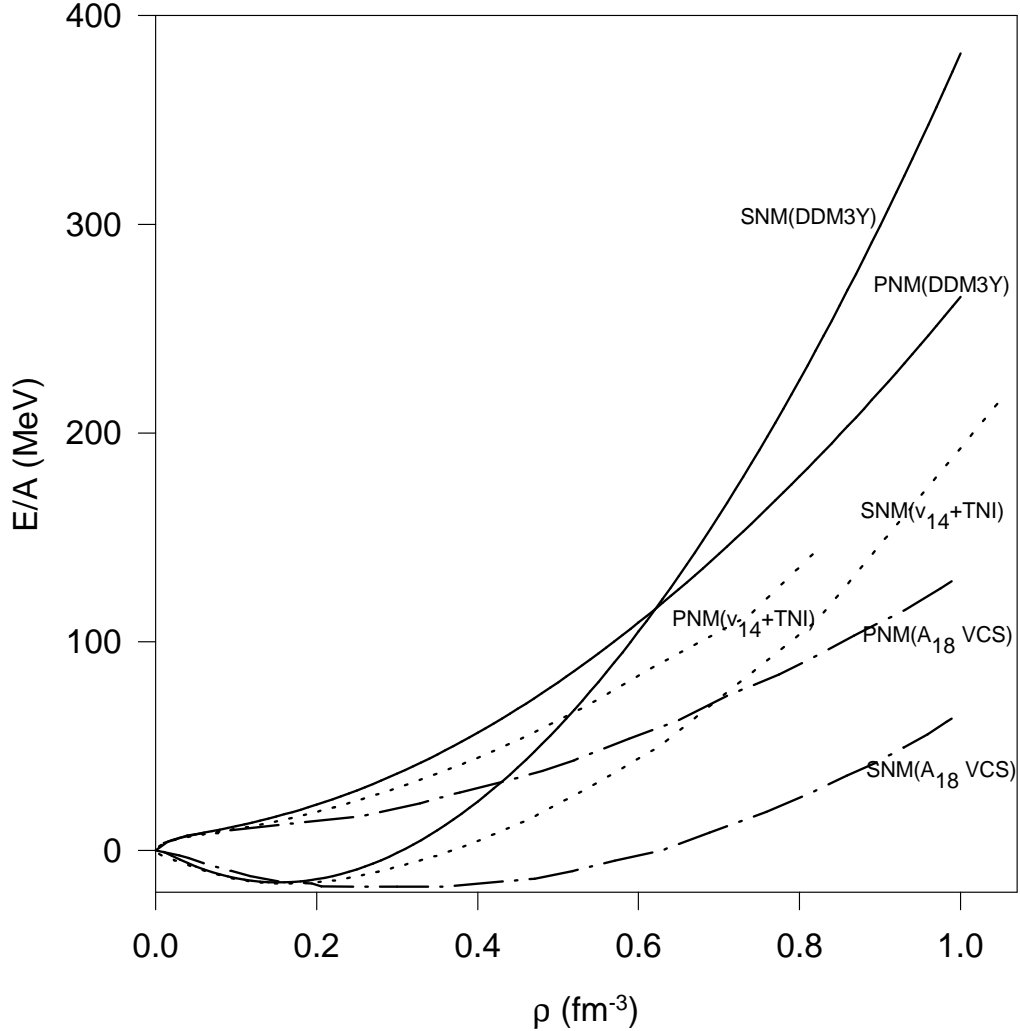,height=14cm,width=14cm}}
\caption
{The energy per nucleon $\epsilon$ = E/A of SNM (spin and isospin symmetric nuclear matter) and PNM (pure neutron matter) as a function of $\rho$. The continuous lines represent curves for the present calculations using saturation energy per nucleon of -15.26 MeV whereas the dotted lines represent the same using $v_{14}+TNI$ interaction [22] and the dash-dotted lines represent the same for the A18 model using variational chain summation (VCS) [24].}
\label{fig1}
\end{figure}

Fig.-2 presents the plots of the energy per nucleon $\epsilon$ of nuclear matter with different isospin asymmetry X as a function of $\rho$ for the present calculations. 

\begin{figure}[htbp]
\eject\centerline{\epsfig{file=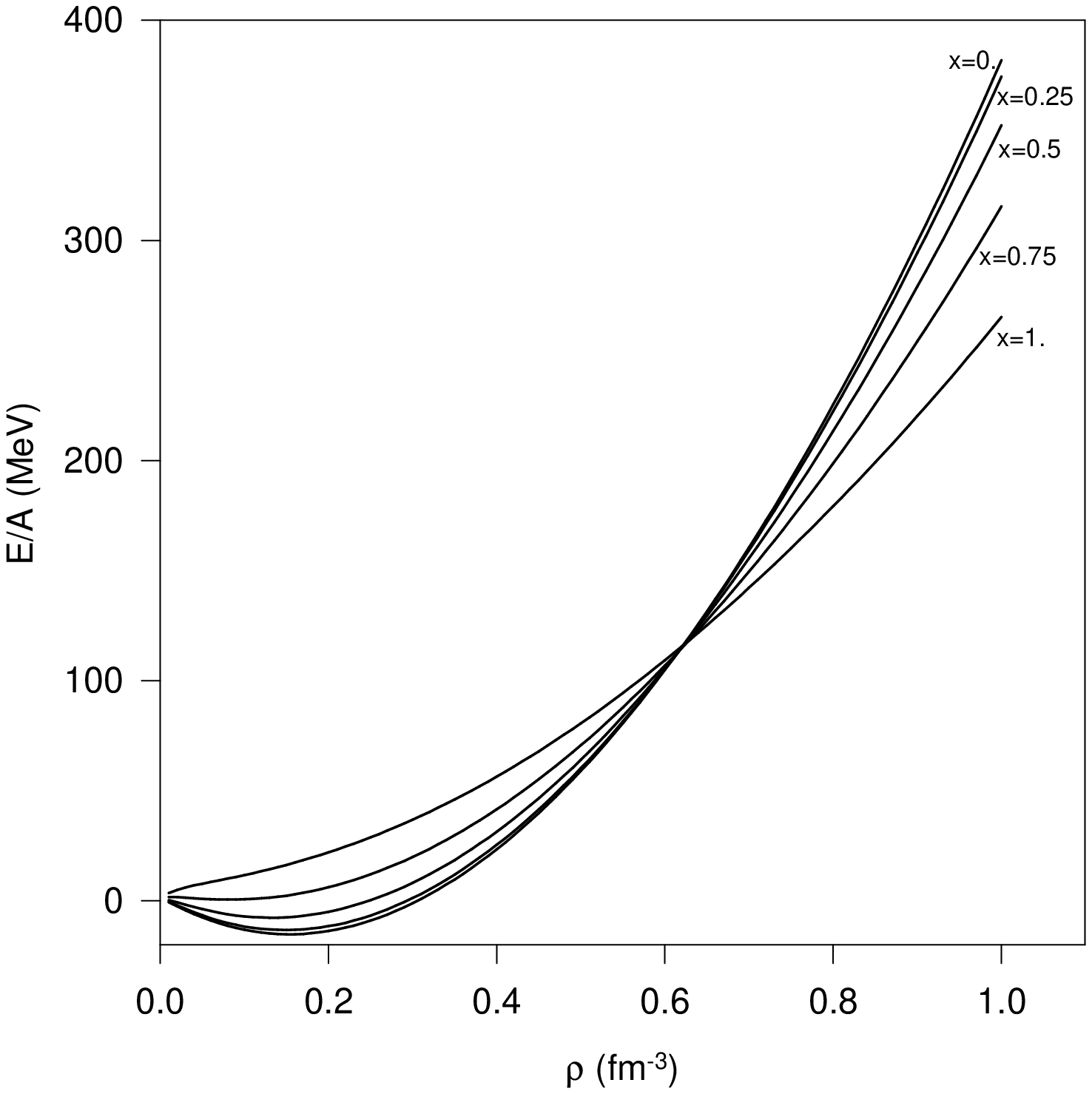,height=14cm,width=14cm}}
\caption
{The energy per nucleon $\epsilon$ = E/A of nuclear matter with different isospin asymmetry X as a function of $\rho$ for the present calculations.}
\label{fig2}
\end{figure}

The pressure $P$ of SNM and PNM are plotted in Fig.-3 as a function of $\rho$. The continuous lines represent the present calculations whereas the dotted lines represent the same using $v_{14}+TNI$ interaction \cite{Fr81}.

\begin{figure}[htbp]
\eject\centerline{\epsfig{file=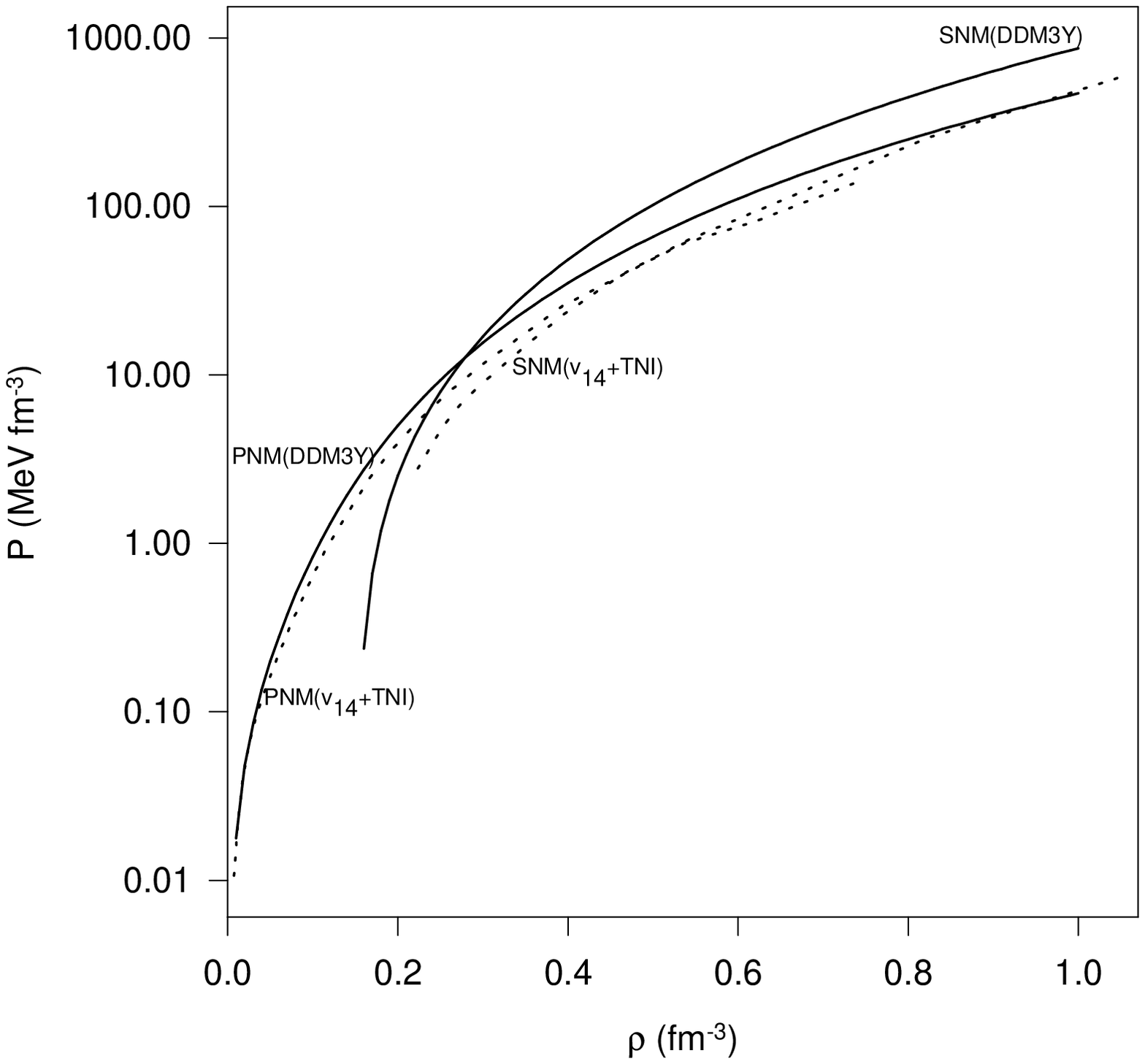,height=14cm,width=14cm}}
\caption
{The pressure $P$ of SNM (spin and isospin symmetric nuclear matter) and PNM (pure neutron matter) as a function of $\rho$. Continuous lines represent the present calculations whereas dotted lines represent the same using $v_{14}+TNI$ interaction [22].}
\label{fig3}
\end{figure}

In Fig.-4 the velocity of sound $v_s$ in SNM and PNM and the energy density $\varepsilon$ of SNM and PNM for the present calculations are plotted as functions of nuleonic density $\rho$. The continuous lines represent the velocity of sound in units of $10^{-2}c$ whereas the dotted lines represent energy density in $MeV fm^{-3}$.

\begin{figure}[htbp]
\eject\centerline{\epsfig{file=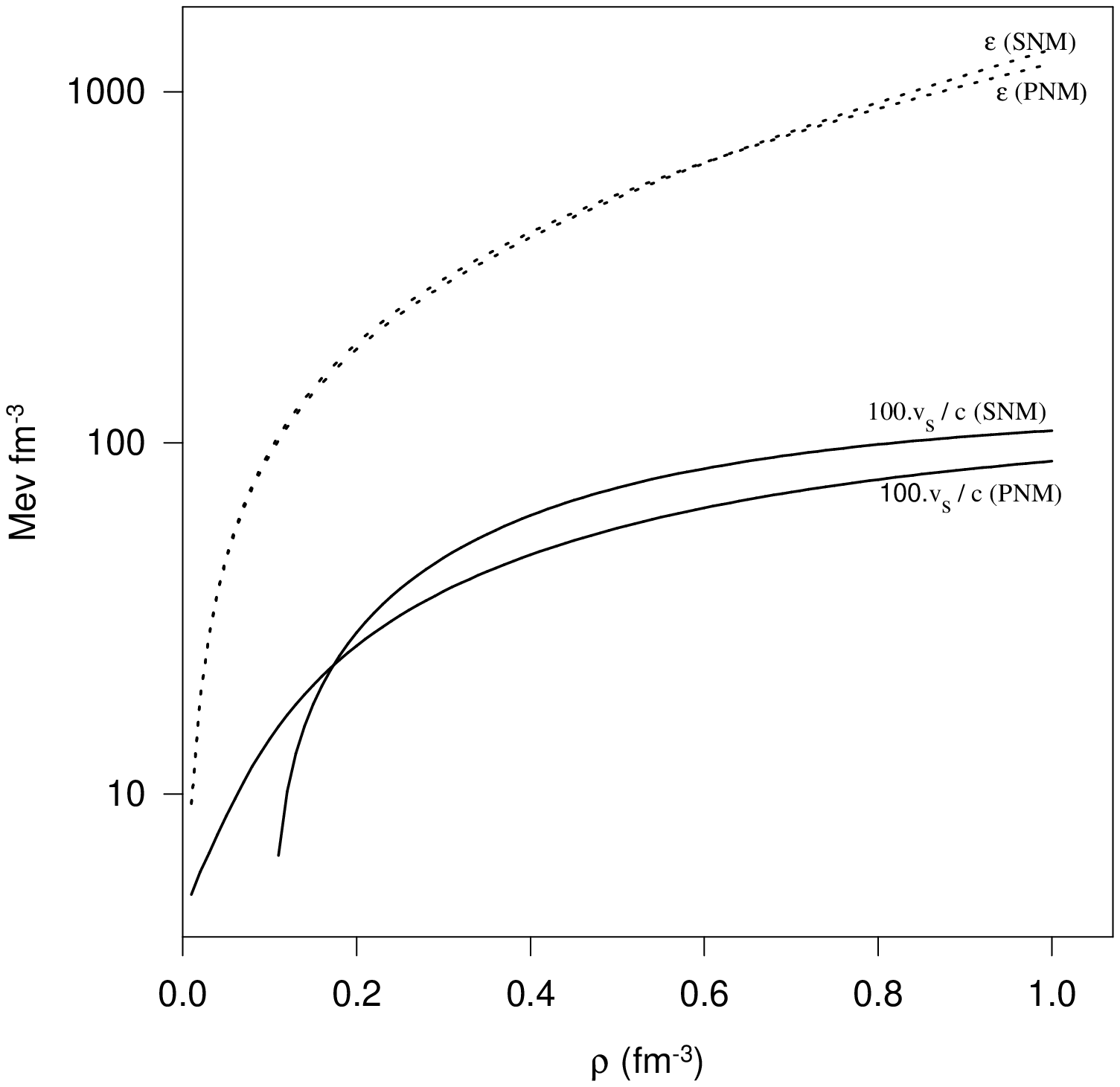,height=14cm,width=14cm}}
\caption
{The velocity of sound $v_s$ in SNM (spin and isospin symmetric nuclear matter) and PNM (pure neutron matter) and the energy density $\varepsilon$ of SNM and PNM as functions of nuleonic density $\rho$ for the present calculations. The continuous lines represent the velocity of sound in units of $10^{-2}c$ whereas the dotted lines represent energy density in $MeV fm^{-3}$.}
\label{fig4}
\end{figure}

      The theoretical estimate $K_0$ of the incompressibility of infinite SNM obtained from present approach using DDM3Y is about $290 MeV$. The theoretical estimate of $K_0$ from the refractive $\alpha$-nucleus scattering is about 240 MeV-270 MeV \cite{Kh97} and that by infinite nuclear matter model (INM) \cite{Sa99} claims a well defined and stable value of $K_0=288\pm20$ MeV and present theoretical estimate is in reasonably close agreement with the value obtained by INM which rules out any values lower than 200 MeV. Present estimate for the incompressibility $K_0$ of the infinite SNM is in good agreement with the experimental value of $K_0=300\pm25$ MeV obtained from the giant monopole resonance (GMR) \cite{Sh88} and with the the recent experimental determination of $K_0$ based upon the production of hard photons in heavy ion collisions which led to the experimental estimate of $K_0=290\pm50$ MeV \cite{Sc96}. However, the experimental values of $K_0$ extracted from the isoscalar giant dipole resonance (ISGDR) are claimed to be smaller \cite{Ga04}. The general theoretical observation by Colo' et al. is that the non-relativistic \cite{Co04} and the relativistic \cite{CG04} mean field models predict for the $K_0$ values which are significantly different from one another, namely $\approx$ 220-235 MeV and $\approx$ 250-270 MeV respectively. Considering the uncertainties in the extractions of $\epsilon_0$ \cite{Ch05} and $\rho_{0}$ values from the experimental masses and electron scattering, present non-relativistic mean field model estimate for the nuclear incompressibility $K_0$ for SNM using DDM3Y interaction is rather close to the earlier theoretical estimates obtained using relativistic mean field models. The density dependence parameter $\beta=1.668 fm^2$, that has the dimension of cross section, can be interpreted as the isospin averaged effective nucleon-nucleon interaction cross section in ground state symmetric nuclear medium. For a nucleon in ground state nuclear matter $k_F\approx$ 1.3 $fm^{-1}$ and $q_0 \sim \hbar k_F c \approx$ 260 MeV and the present result for the 'in medium' effective cross section is reasonably close to the value obtained from a rigorous Dirac-Brueckner-Hartree-Fock \cite{Sa06} calculations corresponding to such $k_F$ and $q_0$ values which is $\approx$ 12 mb. Using the value of the density dependence parameter $\beta=1.668 fm^2$ corresponding to the standard value of the parameter $n=2/3$ along with the nucleonic density of $0.1533 fm^{-3}$, the value obtained for the nuclear mean free path $\lambda$ is about $4 fm$ which is in excellent agreement \cite{Si83} with other theoretical estimates. 

\section{Summary and conclusions}
\label{section5}

      In summary, we conclude that the present EOS is obtained using the isoscalar and Lane that is the isovector  part of M3Y effective NN interaction. This interaction was derived by fitting its matrix elements in an oscillator basis to those elements of the G-matrix obtained with the Reid-Elliot soft-core NN interaction and has a profound theoretical standing. The value obtained for the nuclear mean free path is in excellent agreement \cite{Si83} with other theoretical estimates. The present theoretical estimate of nuclear incompressibility for SNM is in reasonably close agreement with other theoretical estimates obtained by INM \cite{Sa99} model, using the Seyler-Blanchard interaction \cite{Ba90} or the relativistic Brueckner-Hartree-Fock (RBHF) theory \cite{Br90}. This value is also in good agreement with the experimental estimates from GMR \cite{Sh88} as well as determination based upon the production of hard photons in heavy ion collisions \cite{Sc96}. The EOS for SNM and PNM are similar to those obtained by B. Friedman and V.R. Pandharipande using $v_{14}+TNI$ interaction \cite{Fr81} and the RBHF theory. The EOS for the isospin asymmetric nuclear matter can be applied to study the cold compact steller objects such as neutron stars. 




\begin{thebibliography}{9}

\bibitem{Ba80} J.P. Blaizot, Phys. Rep. 65,  171  (1980). 

\bibitem{Sa89} C. Samanta, D. Bandyopadhyay and J.N. De, Phys. Lett. B 217, 381 (1989). 

\bibitem{Be88} G.F. Bertsch and S. Das Gupta, Phys. Rep. 160, 189  (1988).

\bibitem{La62} A.M. Lane, Nucl. Phys. 35, 676 (1962). 

\bibitem{Sa83} G.R. Satchler, Int. series of monographs on Physics, Oxford University Press, Direct Nuclear reactions, 470 (1983).

\bibitem{Be77} G. Bertsch, J. Borysowicz, H. McManus, W.G. Love, Nucl. Phys. A 284, 399 (1977).

\bibitem{Sa79} G.R. Satchler and W.G. Love, Phys. Rep. 55, 183 (1979). 

\bibitem{Ko84}  A.M. Kobos, B.A. Brown, R. Lindsay and G.R. Satchler, Nucl. Phys. A 425, 205 (1984). 

\bibitem{Gi87} H.J. Gils, Nucl. Phys. A 473, 111 (1987).

\bibitem{Gu05} D. Gupta and D.N. Basu, Nucl. Phys. A 748, 402 (2005).

\bibitem{BCS05} D.N. Basu, P. Roy Chowdhury and C. Samanta, Phys. Rev. C 72, 051601(R)  (2005).

\bibitem{Sr83} D.K. Srivastava, D.N. Basu and N.K. Ganguly, Phys. Lett. 124B, 6 (1983).

\bibitem{Ba03} D.N. Basu, Phys. Lett. B 566, 90 (2003).

\bibitem{CSB06} P. Roy Chowdhury, C. Samanta and D. N. Basu, Phys. Rev. C 73, 014612 (2006).

\bibitem{Ba05} D.N. Basu, Int. Jour. Mod. Phys. E 14, 739 (2005).

\bibitem{My73} W.D. Myers, Nucl. Phys. A 204, 465 (1973).

\bibitem{Ba90} D. Bandyopadhyay, C. Samanta, S.K. Samaddar and J.N. De, Nucl. Phys. A 511, 1 (1990).

\bibitem{Ch05} P. Roy Chowdhury, C. Samanta and D. N. Basu, Mod. Phys. Letts. A 21, 1605 (2005).

\bibitem{Au03} G. Audi, A.H. Wapstra and C. Thibault, Nucl. Phys. A 729, 337 (2003). 

\bibitem{Sa99} L. Satpathy, V.S. Uma Maheswari and R.C. Nayak, Phys. Rep. 319, 85 (1999). 

\bibitem{Sc96} Y. Schutz et. al., Nucl. Phys. A 599, 97c (1996).

\bibitem{Fr81} B. Friedman and V.R. Pandharipande, Nucl. Phys. A 361, 502 (1981).

\bibitem{Sh88} M.M. Sharma, W.T.A. Borghols, S. Brandenburg, S. Crona, A. van der Woude and M.N. Harakeh, Phys. Rev. C 38, 2562 (1988). 

\bibitem{Ak98} A. Akmal, V.R. Pandharipande and D.G. Ravenhall, Phys. Rev. C 58, 1804 (1998). 

\bibitem{Kh97} Dao T. Khoa, G.R. Satchler and W. von Oertzen, Phys. Rev. C 56, 954 (1997). 

\bibitem{Ga04} U. Garg, Nucl. Phys. A 731, 3 (2004). 

\bibitem{Co04} G. Colo', N. Van Giai, J. Meyer, K. Bennaceur and P. Bonche, Phys. Rev. C 70, 024307 (2004).

\bibitem{CG04} G. Colo' and N. Van Giai, Nucl. Phys. A 731, 15 (2004).

\bibitem{Sa06} F. Sammarruca and P. Krastev, Phys. Rev. C 73, 014001 (2006).

\bibitem{Si83} B. Sinha, Phys. Rev. Lett. 50, 91 (1983).
 
\bibitem{Br90} R. Brockmann and R. Machleidt, Phys. Rev. C 42, 1965 (1990). 

\end{thebibliography}
\end{document}